\documentclass[a4paper,11pt]{article}
\pdfoutput=1 

\usepackage{jheppub} 

\usepackage[T1]{fontenc} 

\usepackage{amsmath}  
\usepackage{physics}  
\usepackage{slashed}  
\usepackage{amssymb}
\usepackage{bbm}
\usepackage[makeroom]{cancel}
\usepackage{empheq}
\usepackage{verbatim}
\usepackage[colorlinks=true, linkcolor=blue, citecolor=blue, urlcolor=blue]{hyperref}

\newcommand{\g}{\gamma}
\newcommand{\p}{\partial}
\newcommand{\lag}{\mathcal{L}}

\newcommand{\e}{\eta}

\newcommand{\Lam}{\Lambda}
\newcommand{\s}{\sigma}
\newcommand{\m}{\mu}
\newcommand{\n}{\nu}
\newcommand{\rh}{\rho}
\newcommand{\mn}{\mu\nu}

\newcommand{\del}{\delta}

\newcommand{\D}{\mathcal{D}}

\newcommand{\zd}{\dot{z}}

\newcommand{\psid}{\dot{\psi}}

\newcommand{\ezd}{\e \cdot \dot{z}}
\newcommand{\kzd}{k \cdot \dot{z}}

\newcommand{\kpsi}{k \cdot \psi}
\newcommand{\kpsid}{k \cdot \dot{\psi}}
\newcommand{\epsi}{\e \cdot \psi}
\newcommand{\epsid}{\e \cdot \dot{\psi}}
\newcommand{\eDtX}{\e \cdot \D_\tau X}
\newcommand{\kDtX}{k \cdot \D_\tau X}

\newcommand{\Xt}{\widetilde{X}}

\title{\boldmath Interactions of a Continuous-Spin Field with a Spin-1/2 Particle}

\author[a,b]{Shayarneel Kundu,}
\emailAdd{skundu4@stanford.edu}
\author[a,b]{Alessandro Russo,}
\emailAdd{arusso00@stanford.edu}
\author[a]{Philip Schuster,}
\emailAdd{schuster@slac.stanford.edu}
\author[a]{and Natalia Toro}
\emailAdd{ntoro@slac.stanford.edu}

\affiliation[a]{SLAC National Accelerator Laboratory, Stanford University, Stanford, CA 94039, USA}
\affiliation [b]{Department of Physics, Stanford University, Stanford, CA, 94305, USA}

\abstract{We introduce a formalism for coupling a bosonic Continuous-Spin field to familiar spin-1/2 matter. To do this, we describe the matter using the supersymmetric worldline formalism. We construct currents that are local functions of worldline kinematics, and respect both the worldline supersymmetry and the conservation condition required for consistent couplings to Abelian CSP fields.  As the spin Casimir $\rho$ of the CSP vanishes, the interactions reduce to that of familiar QED in one case, and to a Yukawa interaction with a spin-1/2 fermion in another case. Our formalism is applicable to computing deviations from QED if the photon is a CSP, thereby enabling a range of phenomenological studies.}

\begin{document} 
\maketitle
\flushbottom

\section{Introduction} \label{1}

The most general type of Lorentz covariant massless particle in 3+1 dimensions is the so-called ``continuous-spin'' particle (CSP), characterized by a non-zero spin Casimir $W^2=-\rho^2$, where $\rho$ is a continuous parameter with units of momentum. For a review on CSPs see  \cite{Wigner:1939cj}. Particles of this type contain an infinite tower of integer-spaced helicity modes that transform into one another under Lorentz boosts. Only when $\rho=0$ are the helicity modes invariant under boosts, recovering the familiar massless states found in theories such as QED. As demonstrated in \cite{Schuster:2013vpr, Schuster:2014hca, Schuster:2023xqa, Schuster:2023jgc}, theories where CSPs interact with matter necessarily behave like familiar gauge theories at energies $\gg \rho$, where these interactions are dominated by a single primary helicity mode. The novel physics associated with the tower of partner helicity modes primarily impacts the deep infrared. This raises the possibility that some or all Standard Model forces are mediated by CSPs with small but non-zero $\rho$.

In this paper, we consider the interactions of a spin-1/2 matter particle with a bosonic CSP field. Building on the work in \cite{Schuster:2023jgc,Schuster:2023xqa}, we describe the matter by a worldline action and focus on Abelian interactions with a single CSP field.  This approach 
has allowed many aspects of interactions to be studied precisely, as was done in \cite{Schuster:2023jgc} for scalar QED, and in \cite{Kundu:2025fsd} for gravity at leading order. However, only spinless matter has been considered to-date.  This work generalizes the approach of \cite{Schuster:2023xqa} to the physically interesting case of spin-1/2 matter. Our results will provide evidence that there is no principled obstruction to CSPs interacting with matter with spin. Our work further rounds out the theoretical development of CSP physics, where the existing literature has pursued several complementary directions.  Free gauge field theories were first developed for bosonic CSPs in \cite{Schuster:2014hca, Schuster:2013pxj} and for fermionic CSPs in \cite{Alkalaev:2018bqe, Metsaev:2017ytk, Bekaert:2015qkt, Singh:1974rc}. Additional field theoretic aspects were considered in \cite{Metsaev:2018moa, Bekaert:2017xin, Metsaev:2017cuz, Bekaert:2010hp, Berends:1985xx, Berends:1984rq, Iverson:1971hq, Iverson:1970kyn, Schuster:2014xja,Buchbinder:2020nxn, Burdik:2019tzg, Buchbinder:2019esz, Alkalaev:2018bqe, Rivelles:2018tpt, Buchbinder:2018yoo, Alkalaev:2017hvj, Khabarov:2017lth, Bekaert:2017khg, Najafizadeh:2017tin, Zinoviev:2017rnj, Metsaev:2018lth, Rivelles:2016rwo, Bekaert:2015qkt, Rivelles:2014fsa, Schuster:2014hca, Font:2013hia, Bekaert:2010hw, Bekaert:2005in, Khan:2004nj, Segal:2001di, deWit:1979sib, Fang:1978wz, Fronsdal:1978rb, Hirata:1977ss, Abbott:1976bb, Singh:1974qz, Singh:1974rc, Chakrabarti:1971rz, Yngvason:1970fy}, supersymmetric generalizations in \cite{Buchbinder:2022msd, Najafizadeh:2021dsm, Najafizadeh:2019mun, Buchbinder:2019sie, Buchbinder:2019kuh, Buchbinder:2019esz}, and theories in curved spacetimes in \cite{Metsaev:2021zdg, Metsaev:2019opn, Metsaev:2017ytk, Metsaev:2016lhs, Buchbinder:2024jpt, Metsaev:2017myp}. Recently, progress has been made in studying CSP amplitudes using spinor-helicity techniques in \cite{Bellazzini:2024dco}, and the light-cone gauge vector superspace approach in \cite{Metsaev:2025qkr}.

In section \ref{2}, we summarize the worldline description of spin-1/2 particles in terms of an underlying theory with local superysmmetry in 0+1 dimensions, developed by Brink et al in \cite{Brink:1976sz, Brink:1976uf}, including couplings to background scalar and vector gauge fields that are equivalent to Yukawa couplings in field theory and to familiar electrodynamics. In Section \ref{3}, we extend this formalism to include interactions with a background CSP field, which reduce smoothly to the familiar Yukawa and electromagnetic interactions in the limit $\rho \rightarrow 0$.

\section{Spin-1/2 Particles in the Worldline Formalism} \label{2}
\subsection{Brief Review}
\label{2.1}
The dynamics of a relativistic particle can be studied in terms of 0+1 dimensional worldline field theory. To describe a massless fermion in this formalism, we follow the approach of \cite{Brink:1976sz, Brink:1976uf}, where the worldline coordinate $z_\mu (\tau)$, which is Grassmann-even (commuting), is supplemented by a spin degree of freedom $\psi_\mu (\tau)$, which is Grassmann-odd (anti-commuting). Here, $\tau$ is the parameter along the worldline.

Following \cite{Brink:1976sz, Brink:1976uf}, we consider the action for a massless fermion,
\begin{equation}
\label{FreeMasslessFermion}
    S = \frac{1}{2} \int \dd{\tau} \left[ \frac{\dot{z}^2}{e} - i \psi \cdot \dot{\psi} - i \chi \left( \frac{\dot{z} \cdot \psi}{e} \right) \right],  
\end{equation}
with dots denoting derivatives with respect to $\tau$. The new non-dynamical fields, $e$ (einbein) and $\chi$ (einbino), are Grassmann-even and Grassmann-odd respectively. Their equations of motion yield first-class constraints,  
\begin{equation}
    \frac{1}{e^2}\bigg( \dot{z}^2 - i \chi \left( \dot{z} \cdot \psi \right) \bigg) = 0, \qquad \qquad \frac{\dot{z} \cdot \psi}{e} = 0. 
\end{equation}
In terms of the canonical momenta for $z_\mu$ and $\psi_\mu$,
\begin{equation}
    p_\m = \frac{\zd_\m}{e} -  \frac{i \chi \psi_\m}{2 e},  \qquad \Pi_\m = \frac{i \psi_\m}{2},
\end{equation}
these constraints are simply 
\begin{equation}
\label{eq: eom}
    p^2 = 0, \qquad p\cdot \psi = 0.
\end{equation}
Upon quantization of the theory, these constraints will become equations that the physical states must satisfy, namely the Klein-Gordon equation (mass-shell condition) and the Dirac equation. 

To quantize the theory, following Dirac's method, we fix gauge first. A convenient gauge choice is $e =$ constant and $\chi=0$ (we will refer to this as the "flat superspace" gauge). Upon promoting $z_\mu$,$\psi_\mu$, $p_\mu$, and $\Pi_\mu$ to operators, Dirac's quantization then yields the canonical (anti)commutation relations, 
\begin{equation}
    [ z_\m ,  p_\n ] = i \e_{\mn} \qquad \{ \psi_\m,  \Pi_\n \} = i \e_{\mn} \implies \{ \psi_\m , \psi_\n \} = 2 \e_{\mn}.
\end{equation}
The anti-commutation relation for $\psi_\m$ is the Dirac Clifford algebra, and so we can expect that the $\psi^{\mu}$ will act on physical states just as the Dirac matrices $\gamma^{\mu}$ act, and that the physical states will transform in spinor representations. In practice, we can identify $\psi_\m = i \g_5 \g_\m$ \footnote{This differs from the conventions in \cite{Brink:1976uf}, which uses $\psi_\m = \g_5\g_\m$ for massive fermions but $\psi_\m = \g_\m$ for massless fermions. The $\g_5$ factors out and does not affect the results.}. 

\subsection{Worldline Supersymmetry and Superfields} 
\label{2.2}
As discussed in \cite{Brink:1976sz, Brink:1976uf, Berezin:1976eg}, the theory above possesses a local $\mathcal{N}=1$ supersymmetry (SUSY) on the worldline coordinates, where the superpartners are $z_\m$ and $\psi_\m$, and $e$ and $\chi$. Concretely, the free action \eqref{FreeMasslessFermion} is invariant under the transformations
\begin{align}
\label{ComponentFieldTransformation1}
    \delta z^\mu &= a \dot{z}^\mu + i \alpha \psi^\mu, \qquad 
    \delta \psi^\mu = a \dot{\psi}^\mu + \frac{\alpha}{e} \left( \dot{z}^\mu + \chi \frac{i \psi^\mu}{2} \right), \\
\label{ComponentFieldTransformation2}
    \delta e &= a \dot{e} + \dot{a} e + i \alpha \chi, \qquad 
    \delta \chi = a \dot{\chi} + \dot{a} \chi + 2 \dot{\alpha}.
\end{align}
Here, $a(\tau)$ and $\alpha(\tau)$ are respectively, the Grassmann-even and Grassmann-odd parameters of generalized translations and supertranslations. These transformations obey the SUSY algebra appropriate to 0+1 dimensions.  To ensure that supersymmetry is maintained in the presence of interactions, it is useful to introduce a superspace where the worldline parameter $\tau$ is augmented by a Grassmann-odd coordinate $\theta$ and the fields appearing in \eqref{FreeMasslessFermion} can be grouped into Grassmann-even superfields
\begin{equation}
\label{Superfields}
    X^\m (\tau, \theta) = z^\m + i \theta \sqrt{e} \, \psi^\m, \qquad \sqrt{\Lambda (\tau, \theta)} = \sqrt{e} + \frac{i \theta \chi}{2} \, .
\end{equation}
The local SUSY and reparametrization transformations \eqref{ComponentFieldTransformation1}--\eqref{ComponentFieldTransformation2} can be recast as general translations of the superspace coordinates $\tau$, and $\theta$:
\begin{align}
\label{TransformationX}
    & \delta (X^\mu) = \left( b \partial_\tau + \beta \partial_\theta  \right) X^\mu , \\
\label{TransformationLam}
    \delta (\sqrt{\Lam}) &= \partial_\tau (b \sqrt{\Lam}) +  \partial_\theta (\beta \sqrt{\Lam}),
\end{align}
where
\begin{equation}
    \quad b = a + i \theta  \frac{\alpha}{\sqrt{e}}, \quad \text{and} \quad \beta = \frac{\alpha}{\sqrt{e}} + i \theta  \frac{\dot{a}}{2}.
\end{equation}  

We refer to any field that transforms as in \eqref{TransformationX} as a weight-0 superfield, and fields that respect the transformation in \eqref{TransformationLam} as a weight-1 superfield. With these definitions, when superfields are multiplied, their weights add. Thus any function of weight-0 superfields is itself a weight-0 superfield, while its product with a weight-1 superfield transforms as a weight-1 superfield. This property, which we will refer to as the multiplication rule, greatly facilitates building SUSY-invariant actions.

Under supertranslations, weight-1 superfields transform as total derivatives, so their integrals over superspace are invariant under supertranslations. In particular, the measure $d\tau d\theta \sqrt{\Lambda}$ is invariant, and superspace actions of the form
\begin{equation}
\label{InvariantMeasure}
    S=\int d\tau d\theta \sqrt{\Lambda} \lag_0(\tau,\theta),
\end{equation}
are both SUSY- and reparametrization-invariant whenever the superspace Lagrangian $\lag_0(\tau,\theta)$ is constructed entirely from weight-0 superfields.

To build actions with non-trivial dynamics in this way, we must also identify covariant derivatives whose action on a weight-0 superfield returns a weight-0 superfield. The covariant derivatives satisfying this requirement are
\begin{equation}
\label{eq: covariant derivatives}
    \D_\theta  \equiv \frac{1}{\sqrt{\Lambda}} \left( \p_\theta + i \theta \p_\tau \right), \qquad \D_\tau \equiv - i \D_\theta^2.
\end{equation}
Note that $\D_\theta$ and $\D_\tau$ yield superfields of opposite Grassmann parity, that these are the \emph{only} two operators linear in derivatives that satisfy the covariance requirerment, and that in the ``flat superspace'' gauge $\D_\tau$ reduces to the familiar time-derivative $\p_\tau/e$. Other spin-1/2 worldline literature, e.g.~\cite{Brink:1976sz, Brink:1976uf, Mondragon:1994fp}, uses derivative operators that differ by powers of $\Lambda$ and/or additive corrections, and are not in fact covariant in the sense used above. 

The free massless fermion action \eqref{FreeMasslessFermion} can be compactly written in superspace as
\begin{equation}
\label{FreeMasslessSuperfieldAction}
    S_{\text{kin}} = - \frac{i}{2} \int \dd{\tau} \dd{\theta} \sqrt{\Lam} \big( \D_\tau X^\m \D_\theta X_\m \big).
\end{equation}
%

\subsection{Interactions with Background Fields} 
\label{2.3}
\noindent \textbf{Vector Interaction:} In the supersymmetric worldline formalism, the interaction term for spin-$1/2$ matter with an abelian vector background field $A_{\mu}$ is
\begin{equation}
\label{VectorInteraction}
     S_{\text{int}}^{\text{V}} = - \frac{i g_\text{V}}{2} \int \dd{\tau} \dd{\theta} \sqrt{\Lam}\, \D_\theta X_\m A^\m (X).
\end{equation}
The coupling \eqref{VectorInteraction} implies a space-time current
\begin{equation}
\label{VectorCurrent}
    J^{\text{V}}_\m (x) = - \frac{i g_\text{V}}{2} \int \dd{\tau} \dd{\theta} \sqrt{\Lam}\D_\theta X_\m (\tau, \theta) \del^{(4)} \left( x - X(\tau, \theta) \right),
\end{equation}
which has support on a delta function integrated over the worldline in superspace. This current obeys a conservation law $\p^\m J^{\text{V}}_\m (x) = 0$, which one can see by using the delta function to switch the spacetime derivative to a derivative in terms of the superfield, and then integrating by parts. 

Note that $A_\m (X)$ is equivalent by Taylor expansion to
\begin{equation}
    A_\m (X) = A_\m \left(z + i \theta \sqrt{e} \psi \right) = A_\m (z) + i \theta \sqrt{e} \psi^\s \partial_\s A_\m (z),
\end{equation}
which is exact because $\theta^2=0$. 
The interaction term can then be expressed in terms of the component fields as
\begin{equation}
\label{VectorComponentFieldInteraction}
    S_{\text{int}}^{\text{V}} = \frac{g_\text{V}}{2} \int \dd{\tau} \bigg( \zd^\m A_\m (z) - i e \psi^\m \psi^\s \p_\s A_\m (z) \bigg).
\end{equation}
This matches the interaction term in \cite{Brink:1976uf}, when we use our "flat superspace" gauge, and further specify to $e = 1$. In the quantum theory, the interaction \eqref{VectorComponentFieldInteraction} will describe a spin-1/2 particle interacting with an ordinary background vector gauge field. 
\\
\\
\noindent \textbf{Scalar:} The interaction of a spin-1/2 particle with a background scalar field is a little more subtle than for a vector gauge field, and is closely related to massive fermions, as first discussed in \cite{Mondragon:1994fp}. Translating their discussion to the notation used above, we introduce an additional weight-0 superfield,
\begin{equation}
    \Xt = \psi_5 + \frac{i \theta z_5}{\sqrt{e}}.
\end{equation}
The component fields $\psi_5$, and $z_5$ are Grassmann-odd and Grassmann-even respectively, thereby making $\Xt$ a Grassmann-odd superfield. The kinetic term corresponding to this superfield is, 
\begin{equation}
    S^{\Xt}_{\text{kin}} = - \frac{1}{2} \int \dd{\tau} \dd{\theta} \sqrt{\Lam} \Xt \D_\theta \Xt = \frac{1}{2} \int \dd{\tau} \bigg( \frac{z_5^2}{e} + i \psi_5 \psid_5 \bigg).
\end{equation}
With this additional superfield, the interaction term can be written as
\begin{equation}
\label{ScalarInteraction}
    S^{\text{S}}_{\text{int}} = g_{\text{S}} \int \dd{\tau} \dd{\theta} \sqrt{\Lam} \Xt \,\Phi(X).
\end{equation}
Like the vector case, the background scalar field is evaluated on the superfield, and this follows from the current,
\begin{equation}
\label{ScalarCurrent}
    J^{\text{S}}(x) = g_{\text{S}} \int \dd{\tau} \dd{\theta} \sqrt{\Lam} \Xt (\tau, \theta) \del^{(4)} \left( x - X \right).
\end{equation}
Unlike the vector case, there is no conservation condition on this current, since the scalar background has no gauge symmetry. In terms of component fields, the interaction term is 
\begin{equation}
\label{ScalarComponentFieldInteraction}
    S_{\text{int}}^{\text{S}} = g_{\text{S}} \int \dd{\tau} \left( i z_5 \Phi(z) - i e \psi_5 \psi_\s \p^\s \Phi(z) + i \frac{\chi}{2} \psi_5 \Phi(z) \right).
\end{equation}
After integrating out $z_5$, this matches the interaction in \cite{Mondragon:1994fp} with $m=0$ and up to redefinitions of coupling constants. A free action for massive fermions is obtained simply by taking a uniform scalar background $\Phi \rightarrow m/g_{\text{s}}$ in \eqref{ScalarInteraction}.  The interactions with additional scalar or vector fields remain unchanged. 

\section{CSP Field Interaction with Spin $1/2$ Particles} \label{3}

Having introduced the supersymmetric worldline formalism for spin-1/2 particles and their interactions with familiar scalar and vector fields, we now turn to the interactions of spin-1/2 particles with CSPs.  We begin this section with a quick review of the $\e$-space gauge theory of bosonic CSP fields, as developed in \cite{Schuster:2014hca,Schuster:2023xqa}. In the $\e$-space formalism, matter currents must satisfy a continuity condition so that interactions does not excite unphysical modes of the CSP field. In addition, for spin-1/2 matter the current must be invariant under the local worldline supersymmetry described above. With these ingredients, we present currents sourced by spin-1/2 matter that generalize the ``temporal'' current of \cite{Schuster:2023xqa} for spinless matter.

\subsection{CSP Field and $\e$-space}
In the $\e$-space formalism introduced in \cite{Schuster:2023xqa}, CSP fields are functions of both a spacetime coordinate $x^\mu$ and an additional 4-vector coordinate $\e^\m$. CSP fields $\Psi(\eta,x)$ are dynamical on the coordinate space $x$, while the $\eta$ coordinate encodes a tower of symmetric tensors. Roughly, this can be understood by expanding $\Psi$ as, $\Psi(\eta,x)=\psi_0(x)+\psi_1^{\mu}(x)\eta_{\mu}+\psi_2^{\mu\nu}(x)\eta_{\mu}\eta_{\nu}+...$, so that the coefficient functions appearing at each order in $\eta$ are symmetric tensors of increasing rank. In this way, $\Psi$ has enough field content to describe the tower of integer helicity modes of a CSP. Integrations over the auxiliary variable $\eta$ are formally defined via analytic continuation, as introduced in \cite{Schuster:2014hca}. For a review of this formalism, the reader is referred to \cite{Schuster:2023xqa}. The general linear inteaction of a CSP field in this formalism takes the form
\begin{equation}
\label{eq:CSP_current_coupling}
    S^{\text{CSP}}_{\text{int}} = \int \dd[4]{x} [\dd[4]{\e}] \del^{'} (\e^2 + 1) J(\e, x) \Psi(\e, x),
\end{equation}
for some current $J(\e, x)$. Compatibility with the gauge symmetry of the $\Psi$ field implies that the current must satisfy the conservation (continuity) condition
\begin{equation}
    \del (\e^2 + 1) \Big( \p_x \cdot \p_\e + \rh \Big) J(\e, x) = 0,
    \label{eq:continuity}
\end{equation}
where $\p_x$ and $\p_\e$ are space-time and $\eta$-space derivatives respectively. We note that if we expand $J(\eta,x)=J_0 + J_1^\mu \eta_\mu + \dots$ and take the $\rho\to0$ limit, the continuity condition \eqref{eq:continuity} is trivially satisfied by any scalar current $J_0$, and requires the usual conservation condition $\p_x\cdot J_1 = 0$ on the vector current. It can also be shown that inserting these monomial currents into \eqref{eq:CSP_current_coupling} and evaluating the $\eta$-integral simply couples $J_0$ to the scalar component of $\Psi$ and $J_1^\mu$ to the vector component of $\Psi$ (the generalization of the above statements to ranks 2 and higher requires additional trace terms relative to the expansions above, as discussed in \cite{Schuster:2023xqa}).  

To describe a consistent coupling of spin-1/2 particles to CSP fields, without exciting unphysical modes of either theory, a current must \emph{both} respect the continuity condition \eqref{eq:continuity} and be invariant under local SUSY transformations. The latter condition can be assured by building currents as superspace integrals of a function of weight-0 superfields and their covariant derivatives.  Our aim here is to explicitly construct such currents.

Before proceeding for general $\rho$, it is useful to recast the scalar and vector interaction terms \eqref{VectorInteraction}  and \eqref{ScalarInteraction}, which describe interactions with $\rho=0$ fields, in $\e$-space.
The scalar current is given simply by the $\eta$-independent current 
\begin{equation}
\label{eq:scalarCurrentEta}
J^{\text{S}}(\eta,x)= J^{\text{S}}(x) = g_{\text{S}} \int \dd{\tau} \dd{\theta} \sqrt{\Lam} \Xt \del^{(4)} \left( x - X \right)
\end{equation} 
from \eqref{ScalarCurrent}, and the vector current is 
\begin{equation}
\label{eq:vecCurrentEta}
J^{\text{V}}(\eta,x) = \sqrt{2} \e^\mu J^{\text{V}}_\mu(x) = - \frac{i g_\text{V}}{\sqrt{2}} \int \dd{\tau} \dd{\theta} \sqrt{\Lam} \left( \eDtX \right) \del^{(4)} \left( x - X\right).
\end{equation}
Each of these manifestly satisfies the continuity condition (again, up to worldline boundary terms in the vector case) and is worldline-SUSY-invariant as required. 

\subsection{CSP Interaction Currents}
The currents \eqref{eq:scalarCurrentEta} and \eqref{eq:vecCurrentEta} are local over the worldline (i.e.~they take the form of a single integral $\int d\tau j(\tau,x)$, where the "current element" $j(\tau,x)$ depends on worldline kinematic variables at $\tau$ but not at other times).  In addition, their current element at $\tau$ is localized in space-time to the point $z^\mu(\tau)$.  By contrast, in \cite{Schuster:2023xqa} it was found that currents for a spinless worldline that satisfy the CSP continuity condition \eqref{eq:continuity} are local over the worldline, which is crucial for quantizing the interacting theory perturbatively. They \emph{are not}, however, point-localized in space-time\footnote{The "temporal currents" of \cite{Schuster:2023xqa} and the spin-1/2 currents we consider here are, instead, localized to a timelike ray. We refer the reader to Sec.~IV of \cite{Schuster:2023xqa} for discussion of this spacetime localization and its compatibility with causal evolution.}.
Such currents are most simply expressed in momentum-space as 
\begin{equation}
J(\eta^\mu,x^\nu) = \int \frac{\dd[4]{k}}{(2 \pi)^4} \dd\tau e^{- i k \cdot (x - z)} j(\e^\m, k^\n, \tau).
\end{equation}
In superspace for a spin-1/2 worldline these generalize to  
\begin{equation}
\label{CSPCurrent}
    J(\e^\m, x^\n) = \int \frac{\dd[4]{k}}{(2 \pi)^4} \dd{\tau} \dd{\theta} \sqrt{\Lam} e^{- i k \cdot (x - X)} j(\e^\m, k^\n, \tau, \theta).
\end{equation}
Here SUSY-invariance is guaranteed if the current element $j$ is a weight-0 superfield, while the spacetime continuity condition \eqref{eq:continuity} is satisfied provided 
\begin{equation}
\label{ContinuityCondition}
    \Big( - i k \cdot \p_\e + \rh \Big) j(\e^\m, k^\n, \tau, \theta) = 0.
\end{equation}
Currents can be found by considering an appropriate series expansion in $\eta$ and using \eqref{ContinuityCondition} to extract a recursion relation.  Here, inspired by the temporal currents of \cite{Schuster:2023xqa}, which are power series in $\frac{\e\cdot\dot z}{k\cdot z}$, we focus here on currents built out of positive powers of $\tilde X$, $\D_\tau X$, and $\frac{\eDtX}{\kDtX}$ that satisfy the continuity condition and reduce in the $\rho\to 0$ limit to the minimal scalar current \eqref{eq:scalarCurrentEta} or the minimal vector current \eqref{eq:vecCurrentEta}.
\\
\\
\textbf{Coupling to a Scalar-Like CSP:} Following the recursive approach described above, we obtain the scalar-like current element 
\begin{equation}
\label{ScalarCurrent}
    j^{\text{S}}(\eta^\mu,k^\nu,\tau,\theta) = \widetilde{X} \exp \left( - i \rh \frac{\eDtX}{\kDtX} \right).
\end{equation}
This trivially satisfies the continuity condition and, in the $\rho\to 0$ limit, recovers the scalar current element $j = \widetilde{X}$ in agreement with \eqref{eq:scalarCurrentEta}.
In the "flat superspace" gauge, the current can be written in terms of component fields as
\begin{equation}
    J^{\text{S}} (\eta^\mu, x^\nu) = g_{\text{S}}^{\text{CSP}} \int d\tau \, \frac{d^4 k}{(2\pi)^4} e^{-i k  (x - z)} e^{-i \rho \frac{\ezd}{\kzd}} 
    \left[ e \psi_5 \left( \kpsi - \rho \left( \epsid  - \frac{(\ezd )( \kpsid)}{(\kzd)} \right) \right) + i \zd_5 \right].
\end{equation}
Here we have used \refeq{CSPCurrent}, and written the current in spacetime, as an inverse Fourier transform.
\\
\\
\textbf{Coupling to a Vector-Like CSP:} Using the same phase factor structure as before \eqref{ScalarCurrent}, but now demanding that we recover \eqref{eq:vecCurrentEta} in the $\rho\to 0$ limit, we find the current
\begin{equation}
\label{VectorCurrent}
    j^{\text{V}} = \sqrt{2} \exp \left( -i \rho \frac{\eDtX}{\kDtX} \right) \left[ \left( \eta \cdot \D_\theta X \right) + \frac{\left(k \cdot \D_\theta X \right)}{i \rho} \left( - i \rho \frac{\eDtX}{\kDtX} - 1 \right) \right].
\end{equation}
It is instructive to consider a power series in $\rh$. In the small-$\rh$ limit, we find that the term proportional to $1/\rh$ is a total derivative, and therefore does not contribute to the action. The leading non-trivial contributions to the series expansion are
\begin{equation}
    j^{\text{V}}(\eta,k,\tau,\theta) = \sqrt{2} \left( \eta \cdot \D_\theta X \right) + i \rho \sqrt{2} \left[ \frac{\left( \eDtX \right) \left(k \cdot \D_\theta X \right)  \left( \eDtX \right)}{\left( \kDtX \right) \left( \kDtX \right)} - \frac{\left( \eDtX \right) \left( \eta \cdot \D_\theta X \right)}{\left( \kDtX \right)} \right]. 
\end{equation}
When setting $\rh = 0$, we recover $j \to \sqrt{2} \eta \cdot \D_\theta X$, which corresponds to the current used for the vector coupling in Section~\ref{2.2}. The second term represents the leading correction to the interaction at small but non-zero $\rho$. 

The structure of the current here is more complex than in the scalar case of \eqref{ScalarCurrent}. Notably, we see terms proportional to $\D_\theta X_\mu$, which are Grassmann-odd. This reflects the contribution arising from the coupling of the fermionic spin degrees of freedom to the spin structure of the CSP field. In the "flat superspace" gauge, the component form of the space-time current is
\begin{align}
     J^{\text{V}} (\eta^\mu, x^\nu) &= - \frac{g_{\text{V}}^{\text{CSP}}}{\sqrt{2}} \int \dd{\tau}  \frac{\dd[4]{k}}{(2 \pi)^4} e^{-ik(x-z)}e^{-i\rho\frac{\ezd}{\kzd}} \, \times \nonumber \\
     &\quad \Bigg( -\frac{(\kzd)}{\rho} - i e (\kpsi)(\epsi) + i e\rho \, \frac{4 \, \eta^{[\mu} k^{\nu]} \eta^{[\sigma} k^{\gamma]}}{(\kzd)^3} \, \Dot{\psi}_\mu\Dot{z}_\nu \psi_\sigma\Dot{z}_\gamma
     \Bigg).
\end{align}

\section{Conclusion}
\label{conclusion} 
In this paper, we have generalized the work of \cite{Schuster:2023xqa} by constructing appropriately conserved currents for  spin-1/2 particles coupled to CSP fields.
This is achieved starting from the treatment of spin-1/2 particles using a 0+1 dimensional supersymmetric worldline formalism. We have constructed interaction currents that respect worldline supersymmetry and satisfy the CSP continuity condition in $\eta$-space. These currents are equivalent to familiar QED and Yukawa interactions in the $\rho \rightarrow 0$ limit, making them interesting to consider in the context of CSP phenomenology.  In particular, our formalism provides a pathway for extending the thermodynamics study of \cite{Schuster:2024wjc}, to analyze deviations from QED in spin-1/2 matter systems for $\rho \neq 0$ \cite{Schuster:2023jgc, Reilly:2025lnm}, and to explore spin-dependent gravitational wave observables as studied in \cite{Kundu:2025fsd}. 

Future directions include extending our formalism to $\mathcal{N}=2$ worldline supersymmetry, which would presumably facilitate the description of a standard $h=1$ photon interacting with a CSP field. This would be interesting because the results of \cite{Bellazzini:2024dco} encountered obstructions to 3-particle on-shell amplitudes with two CSPs and standard $h=1,2$ massless particles. In the formalism of this paper, scattering interactions are described with 3-particle (off-shell) ``vertex operators'' defined on the matter worldline, and using these, one can compute on-shell amplitudes with four or more external particle, as was done in \cite{Schuster:2023jgc} for CSP QED with scalar matter.


\acknowledgments
The authors are supported by the U.S. Department of Energy under contract number DE-AC02-76SF00515.

\appendix


\bibliographystyle{unsrtnat}
\bibliography{refs.bib}

\begin{thebibliography}{64}
\providecommand{\natexlab}[1]{#1}
\providecommand{\url}[1]{\texttt{#1}}
\expandafter\ifx\csname urlstyle\endcsname\relax
  \providecommand{\doi}[1]{doi: #1}\else
  \providecommand{\doi}{doi: \begingroup \urlstyle{rm}\Url}\fi

\bibitem[Wigner(1939)]{Wigner:1939cj}
Eugene~P. Wigner.
\newblock {On Unitary Representations of the Inhomogeneous Lorentz Group}.
\newblock \emph{Annals Math.}, 40:\penalty0 149--204, 1939.
\newblock \doi{10.2307/1968551}.

\bibitem[Schuster and Toro(2013{\natexlab{a}})]{Schuster:2013vpr}
Philip Schuster and Natalia Toro.
\newblock {On the Theory of Continuous-Spin Particles: Helicity Correspondence in Radiation and Forces}.
\newblock \emph{JHEP}, 09:\penalty0 105, 2013{\natexlab{a}}.
\newblock \doi{10.1007/JHEP09(2013)105}.

\bibitem[Schuster and Toro(2015{\natexlab{a}})]{Schuster:2014hca}
Philip Schuster and Natalia Toro.
\newblock {Continuous-spin particle field theory with helicity correspondence}.
\newblock \emph{Phys. Rev. D}, 91:\penalty0 025023, 2015{\natexlab{a}}.
\newblock \doi{10.1103/PhysRevD.91.025023}.

\bibitem[Schuster et~al.(2023)Schuster, Toro, and Zhou]{Schuster:2023xqa}
Philip Schuster, Natalia Toro, and Kevin Zhou.
\newblock {Interactions of Particles with ''Continuous Spin'' Fields}.
\newblock \emph{JHEP}, 04:\penalty0 010, 2023.
\newblock \doi{10.1007/JHEP04(2023)010}.

\bibitem[Schuster and Toro(2024)]{Schuster:2023jgc}
Philip Schuster and Natalia Toro.
\newblock {Quantum electrodynamics mediated by a photon with continuous spin}.
\newblock \emph{Phys. Rev. D}, 109\penalty0 (9):\penalty0 096008, 2024.
\newblock \doi{10.1103/PhysRevD.109.096008}.

\bibitem[Kundu et~al.(2025)Kundu, Schuster, and Toro]{Kundu:2025fsd}
Shayarneel Kundu, Philip Schuster, and Natalia Toro.
\newblock {A First Look at ''Continuous Spin'' Gravity -- Time Delay Signatures}.
\newblock arXiv:2503.03817 [gr-qc], March 2025.

\bibitem[Schuster and Toro(2013{\natexlab{b}})]{Schuster:2013pxj}
Philip Schuster and Natalia Toro.
\newblock {On the Theory of Continuous-Spin Particles: Wavefunctions and Soft-Factor Scattering Amplitudes}.
\newblock \emph{JHEP}, 09:\penalty0 104, 2013{\natexlab{b}}.
\newblock \doi{10.1007/JHEP09(2013)104}.

\bibitem[Alkalaev et~al.(2018)Alkalaev, Chekmenev, and Grigoriev]{Alkalaev:2018bqe}
Konstantin Alkalaev, Alexander Chekmenev, and Maxim Grigoriev.
\newblock {Unified formulation for helicity and continuous spin fermionic fields}.
\newblock \emph{JHEP}, 11:\penalty0 050, 2018.
\newblock \doi{10.1007/JHEP11(2018)050}.

\bibitem[Metsaev(2017{\natexlab{a}})]{Metsaev:2017ytk}
R.~R. Metsaev.
\newblock {Fermionic continuous spin gauge field in (A)dS space}.
\newblock \emph{Phys. Lett. B}, 773:\penalty0 135--141, 2017{\natexlab{a}}.
\newblock \doi{10.1016/j.physletb.2017.08.020}.

\bibitem[Bekaert et~al.(2016)Bekaert, Najafizadeh, and Setare]{Bekaert:2015qkt}
X.~Bekaert, M.~Najafizadeh, and M.~R. Setare.
\newblock {A gauge field theory of fermionic Continuous-Spin Particles}.
\newblock \emph{Phys. Lett. B}, 760:\penalty0 320--323, 2016.
\newblock \doi{10.1016/j.physletb.2016.07.005}.

\bibitem[Singh and Hagen(1974{\natexlab{a}})]{Singh:1974rc}
L.~P.~S. Singh and C.~R. Hagen.
\newblock {Lagrangian formulation for arbitrary spin. 2. The fermion case}.
\newblock \emph{Phys. Rev. D}, 9:\penalty0 910--920, 1974{\natexlab{a}}.
\newblock \doi{10.1103/PhysRevD.9.910}.

\bibitem[Metsaev(2018{\natexlab{a}})]{Metsaev:2018moa}
R.~R. Metsaev.
\newblock {Cubic interaction vertices for massive/massless continuous-spin fields and arbitrary spin fields}.
\newblock \emph{JHEP}, 12:\penalty0 055, 2018{\natexlab{a}}.
\newblock \doi{10.1007/JHEP12(2018)055}.

\bibitem[Bekaert et~al.(2017)Bekaert, Mourad, and Najafizadeh]{Bekaert:2017xin}
Xavier Bekaert, Jihad Mourad, and Mojtaba Najafizadeh.
\newblock {Continuous-spin field propagator and interaction with matter}.
\newblock \emph{JHEP}, 11:\penalty0 113, 2017.
\newblock \doi{10.1007/JHEP11(2017)113}.

\bibitem[Metsaev(2017{\natexlab{b}})]{Metsaev:2017cuz}
R.~R. Metsaev.
\newblock {Cubic interaction vertices for continuous-spin fields and arbitrary spin massive fields}.
\newblock \emph{JHEP}, 11:\penalty0 197, 2017{\natexlab{b}}.
\newblock \doi{10.1007/JHEP11(2017)197}.

\bibitem[Bekaert et~al.(2010)Bekaert, Boulanger, and Leclercq]{Bekaert:2010hp}
Xavier Bekaert, Nicolas Boulanger, and Serge Leclercq.
\newblock {Strong obstruction of the Berends-Burgers-van Dam spin-3 vertex}.
\newblock \emph{J. Phys. A}, 43:\penalty0 185401, 2010.
\newblock \doi{10.1088/1751-8113/43/18/185401}.

\bibitem[Berends et~al.(1986)Berends, Burgers, and van Dam]{Berends:1985xx}
Frits~A. Berends, G.~J.~H. Burgers, and H.~van Dam.
\newblock {Explicit construction of conserved currents for massless fields of arbitrary spin}.
\newblock \emph{Nucl. Phys. B}, 271:\penalty0 429--441, 1986.
\newblock \doi{10.1016/S0550-3213(86)80019-0}.

\bibitem[Berends et~al.(1985)Berends, Burgers, and van Dam]{Berends:1984rq}
Frits~A. Berends, G.~J.~H. Burgers, and H.~van Dam.
\newblock {On the Theoretical Problems in Constructing Interactions Involving Higher Spin Massless Particles}.
\newblock \emph{Nucl. Phys. B}, 260:\penalty0 295--322, 1985.
\newblock \doi{10.1016/0550-3213(85)90074-4}.

\bibitem[Iverson and Mack(1971)]{Iverson:1971hq}
G.~J. Iverson and G.~Mack.
\newblock {Quantum fields and interactions of massless particles - the continuous spin case}.
\newblock \emph{Annals Phys.}, 64:\penalty0 211--253, 1971.
\newblock \doi{10.1016/0003-4916(71)90284-3}.

\bibitem[Iverson and Mack(1970)]{Iverson:1970kyn}
G.~J. Iverson and G.~Mack.
\newblock {Theory of weak interactions with *continuous-spin* neutrinos}.
\newblock \emph{Phys. Rev. D}, 2:\penalty0 2326--2333, 1970.
\newblock \doi{10.1103/PhysRevD.2.2326}.

\bibitem[Schuster and Toro(2015{\natexlab{b}})]{Schuster:2014xja}
Philip Schuster and Natalia Toro.
\newblock {A new class of particle in 2 + 1 dimensions}.
\newblock \emph{Phys. Lett. B}, 743:\penalty0 224--227, 2015{\natexlab{b}}.
\newblock \doi{10.1016/j.physletb.2015.02.050}.

\bibitem[Buchbinder et~al.(2020{\natexlab{a}})Buchbinder, Fedoruk, Isaev, and Krykhtin]{Buchbinder:2020nxn}
I.~L. Buchbinder, S.~Fedoruk, A.~P. Isaev, and V.~A. Krykhtin.
\newblock {Towards Lagrangian construction for infinite half-integer spin field}.
\newblock \emph{Nucl. Phys. B}, 958:\penalty0 115114, 2020{\natexlab{a}}.
\newblock \doi{10.1016/j.nuclphysb.2020.115114}.

\bibitem[Burd\'\i{}k et~al.(2020)Burd\'\i{}k, Pandey, and Reshetnyak]{Burdik:2019tzg}
\v{C}. Burd\'\i{}k, V.~K. Pandey, and A.~Reshetnyak.
\newblock {BRST\textendash{}BFV and BRST\textendash{}BV descriptions for bosonic fields with continuous spin on $R^{1,d-1}$}.
\newblock \emph{Int. J. Mod. Phys. A}, 35\penalty0 (26):\penalty0 2050154, 2020.
\newblock \doi{10.1142/S0217751X20501547}.

\bibitem[Buchbinder et~al.(2019{\natexlab{a}})Buchbinder, Gates, and Koutrolikos]{Buchbinder:2019esz}
I.~L. Buchbinder, S.~James Gates, and K.~Koutrolikos.
\newblock {Superfield continuous spin equations of motion}.
\newblock \emph{Phys. Lett. B}, 793:\penalty0 445--450, 2019{\natexlab{a}}.
\newblock \doi{10.1016/j.physletb.2019.05.015}.

\bibitem[Rivelles(2018)]{Rivelles:2018tpt}
Victor~O. Rivelles.
\newblock {A Gauge Field Theory for Continuous Spin Tachyons}.
\newblock arXiv:1807.01812 [hep-th], July 2018.

\bibitem[Buchbinder et~al.(2018)Buchbinder, Krykhtin, and Takata]{Buchbinder:2018yoo}
I.~L. Buchbinder, V.~A. Krykhtin, and H.~Takata.
\newblock {BRST approach to Lagrangian construction for bosonic continuous spin field}.
\newblock \emph{Phys. Lett. B}, 785:\penalty0 315--319, 2018.
\newblock \doi{10.1016/j.physletb.2018.07.070}.

\bibitem[Alkalaev and Grigoriev(2018)]{Alkalaev:2017hvj}
Konstantin~B. Alkalaev and Maxim~A. Grigoriev.
\newblock {Continuous spin fields of mixed-symmetry type}.
\newblock \emph{JHEP}, 03:\penalty0 030, 2018.
\newblock \doi{10.1007/JHEP03(2018)030}.

\bibitem[Khabarov and Zinoviev(2018)]{Khabarov:2017lth}
M.~V. Khabarov and Yu.~M. Zinoviev.
\newblock {Infinite (continuous) spin fields in the frame-like formalism}.
\newblock \emph{Nucl. Phys. B}, 928:\penalty0 182--216, 2018.
\newblock \doi{10.1016/j.nuclphysb.2018.01.016}.

\bibitem[Bekaert and Skvortsov(2017)]{Bekaert:2017khg}
Xavier Bekaert and Evgeny~D. Skvortsov.
\newblock {Elementary particles with continuous spin}.
\newblock \emph{Int. J. Mod. Phys. A}, 32\penalty0 (23n24):\penalty0 1730019, 2017.
\newblock \doi{10.1142/S0217751X17300198}.

\bibitem[Najafizadeh(2018)]{Najafizadeh:2017tin}
Mojtaba Najafizadeh.
\newblock {Modified Wigner equations and continuous spin gauge field}.
\newblock \emph{Phys. Rev. D}, 97\penalty0 (6):\penalty0 065009, 2018.
\newblock \doi{10.1103/PhysRevD.97.065009}.

\bibitem[Zinoviev(2017)]{Zinoviev:2017rnj}
Yu.~M. Zinoviev.
\newblock {Infinite spin fields in d = 3 and beyond}.
\newblock \emph{Universe}, 3\penalty0 (3):\penalty0 63, 2017.
\newblock \doi{10.3390/universe3030063}.

\bibitem[Metsaev(2018{\natexlab{b}})]{Metsaev:2018lth}
R.~R. Metsaev.
\newblock {BRST-BV approach to continuous-spin field}.
\newblock \emph{Phys. Lett. B}, 781:\penalty0 568--573, 2018{\natexlab{b}}.
\newblock \doi{10.1016/j.physletb.2018.04.038}.

\bibitem[Rivelles(2017)]{Rivelles:2016rwo}
Victor~O. Rivelles.
\newblock {Remarks on a Gauge Theory for Continuous Spin Particles}.
\newblock \emph{Eur. Phys. J. C}, 77\penalty0 (7):\penalty0 433, 2017.
\newblock \doi{10.1140/epjc/s10052-017-4927-1}.

\bibitem[Rivelles(2015)]{Rivelles:2014fsa}
Victor~O. Rivelles.
\newblock {Gauge Theory Formulations for Continuous and Higher Spin Fields}.
\newblock \emph{Phys. Rev. D}, 91\penalty0 (12):\penalty0 125035, 2015.
\newblock \doi{10.1103/PhysRevD.91.125035}.

\bibitem[Font et~al.(2014)Font, Quevedo, and Theisen]{Font:2013hia}
Anamaria Font, Fernando Quevedo, and Stefan Theisen.
\newblock {A comment on continuous spin representations of the Poincare group and perturbative string theory}.
\newblock \emph{Fortsch. Phys.}, 62:\penalty0 975--980, 2014.
\newblock \doi{10.1002/prop.201400067}.

\bibitem[Bekaert et~al.(2012)Bekaert, Boulanger, and Sundell]{Bekaert:2010hw}
Xavier Bekaert, Nicolas Boulanger, and Per Sundell.
\newblock {How higher-spin gravity surpasses the spin two barrier: no-go theorems versus yes-go examples}.
\newblock \emph{Rev. Mod. Phys.}, 84:\penalty0 987--1009, 2012.
\newblock \doi{10.1103/RevModPhys.84.987}.

\bibitem[Bekaert and Mourad(2006)]{Bekaert:2005in}
X.~Bekaert and J.~Mourad.
\newblock {The Continuous spin limit of higher spin field equations}.
\newblock \emph{JHEP}, 01:\penalty0 115, 2006.
\newblock \doi{10.1088/1126-6708/2006/01/115}.

\bibitem[Khan and Ramond(2005)]{Khan:2004nj}
Abu~M. Khan and Pierre Ramond.
\newblock {Continuous spin representations from group contraction}.
\newblock \emph{J. Math. Phys.}, 46:\penalty0 053515, 2005.
\newblock \doi{10.1063/1.1897663}.
\newblock [Erratum: J.Math.Phys. 46, 079901 (2005)].

\bibitem[Segal(2003)]{Segal:2001di}
Arkady~Yu. Segal.
\newblock {Point particle in general background fields vsersus gauge theories of traceless symmetric tensors}.
\newblock \emph{Int. J. Mod. Phys. A}, 18:\penalty0 4999--5021, 2003.
\newblock \doi{10.1142/S0217751X03015830}.

\bibitem[de~Wit and Freedman(1980)]{deWit:1979sib}
Bernard de~Wit and Daniel~Z. Freedman.
\newblock {Systematics of Higher Spin Gauge Fields}.
\newblock \emph{Phys. Rev. D}, 21:\penalty0 358, 1980.
\newblock \doi{10.1103/PhysRevD.21.358}.

\bibitem[Fang and Fronsdal(1978)]{Fang:1978wz}
J.~Fang and C.~Fronsdal.
\newblock {Massless Fields with Half Integral Spin}.
\newblock \emph{Phys. Rev. D}, 18:\penalty0 3630, 1978.
\newblock \doi{10.1103/PhysRevD.18.3630}.

\bibitem[Fronsdal(1978)]{Fronsdal:1978rb}
Christian Fronsdal.
\newblock {Massless Fields with Integer Spin}.
\newblock \emph{Phys. Rev. D}, 18:\penalty0 3624, 1978.
\newblock \doi{10.1103/PhysRevD.18.3624}.

\bibitem[Hirata(1977)]{Hirata:1977ss}
K.~Hirata.
\newblock {Quantization of Massless Fields with Continuous Spin}.
\newblock \emph{Prog. Theor. Phys.}, 58:\penalty0 652--666, 1977.
\newblock \doi{10.1143/PTP.58.652}.

\bibitem[Abbott(1976)]{Abbott:1976bb}
L.~F. Abbott.
\newblock {Massless Particles with Continuous Spin Indices}.
\newblock \emph{Phys. Rev. D}, 13:\penalty0 2291, 1976.
\newblock \doi{10.1103/PhysRevD.13.2291}.

\bibitem[Singh and Hagen(1974{\natexlab{b}})]{Singh:1974qz}
L.~P.~S. Singh and C.~R. Hagen.
\newblock {Lagrangian formulation for arbitrary spin. 1. The boson case}.
\newblock \emph{Phys. Rev. D}, 9:\penalty0 898--909, 1974{\natexlab{b}}.
\newblock \doi{10.1103/PhysRevD.9.898}.

\bibitem[Chakrabarti(1971)]{Chakrabarti:1971rz}
A.~Chakrabarti.
\newblock {Remarks on lightlike continuous spin and spacelike representations of the poincare group}.
\newblock \emph{J. Math. Phys.}, 12:\penalty0 1813--1822, 1971.
\newblock \doi{10.1063/1.1665809}.

\bibitem[Yngvason(1970)]{Yngvason:1970fy}
J.~Yngvason.
\newblock {Zero-mass infinite spin representations of the poincare group and quantum field theory}.
\newblock \emph{Commun. Math. Phys.}, 18:\penalty0 195--203, 1970.
\newblock \doi{10.1007/BF01649432}.

\bibitem[Buchbinder et~al.(2022)Buchbinder, Fedoruk, Isaev, and Krykhtin]{Buchbinder:2022msd}
I.~L. Buchbinder, S.~A. Fedoruk, A.~P. Isaev, and V.~A. Krykhtin.
\newblock {On the off-shell superfield Lagrangian formulation of 4D, N=1 supersymmetric infinite spin theory}.
\newblock \emph{Phys. Lett. B}, 829:\penalty0 137139, 2022.
\newblock \doi{10.1016/j.physletb.2022.137139}.

\bibitem[Najafizadeh(2022)]{Najafizadeh:2021dsm}
Mojtaba Najafizadeh.
\newblock {Off-shell supersymmetric continuous spin gauge theory}.
\newblock \emph{JHEP}, 02:\penalty0 038, 2022.
\newblock \doi{10.1007/JHEP02(2022)038}.

\bibitem[Najafizadeh(2020)]{Najafizadeh:2019mun}
Mojtaba Najafizadeh.
\newblock {Supersymmetric Continuous Spin Gauge Theory}.
\newblock \emph{JHEP}, 03:\penalty0 027, 2020.
\newblock \doi{10.1007/JHEP03(2020)027}.

\bibitem[Buchbinder et~al.(2020{\natexlab{b}})Buchbinder, Isaev, and Fedoruk]{Buchbinder:2019sie}
I.~L. Buchbinder, A.~P. Isaev, and S.~A. Fedoruk.
\newblock {Massless Infinite Spin (Super)particles and Fields}.
\newblock \emph{Proc. Steklov Inst. Math.}, 309\penalty0 (1):\penalty0 46--56, 2020{\natexlab{b}}.
\newblock \doi{10.1134/S0081543820030049}.

\bibitem[Buchbinder et~al.(2019{\natexlab{b}})Buchbinder, Khabarov, Snegirev, and Zinoviev]{Buchbinder:2019kuh}
I.~L. Buchbinder, M.~V. Khabarov, T.~V. Snegirev, and Yu.~M. Zinoviev.
\newblock {Lagrangian formulation for the infinite spin $N$=1 supermultiplets in $d$=4}.
\newblock \emph{Nucl. Phys. B}, 946:\penalty0 114717, 2019{\natexlab{b}}.
\newblock \doi{10.1016/j.nuclphysb.2019.114717}.

\bibitem[Metsaev(2021)]{Metsaev:2021zdg}
R.~R. Metsaev.
\newblock {Mixed-symmetry continuous-spin fields in flat and AdS spaces}.
\newblock \emph{Phys. Lett. B}, 820:\penalty0 136497, 2021.
\newblock \doi{10.1016/j.physletb.2021.136497}.

\bibitem[Metsaev(2019)]{Metsaev:2019opn}
R.~R. Metsaev.
\newblock {Light-cone continuous-spin field in AdS space}.
\newblock \emph{Phys. Lett. B}, 793:\penalty0 134--140, 2019.
\newblock \doi{10.1016/j.physletb.2019.04.041}.

\bibitem[Metsaev(2017{\natexlab{c}})]{Metsaev:2016lhs}
R.~R. Metsaev.
\newblock {Continuous spin gauge field in (A)dS space}.
\newblock \emph{Phys. Lett. B}, 767:\penalty0 458--464, 2017{\natexlab{c}}.
\newblock \doi{10.1016/j.physletb.2017.02.027}.

\bibitem[Buchbinder et~al.(2024)Buchbinder, Fedoruk, Isaev, and Krykhtin]{Buchbinder:2024jpt}
I.~L. Buchbinder, S.~A. Fedoruk, A.~P. Isaev, and V.~A. Krykhtin.
\newblock {BRST construction for infinite spin field on $AdS_4$}.
\newblock \emph{Eur. Phys. J. Plus}, 139\penalty0 (7):\penalty0 621, 2024.
\newblock \doi{10.1140/epjp/s13360-024-05430-6}.

\bibitem[Metsaev(2018{\natexlab{c}})]{Metsaev:2017myp}
R.~R. Metsaev.
\newblock {Continuous-spin mixed-symmetry fields in AdS(5)}.
\newblock \emph{J. Phys. A}, 51\penalty0 (21):\penalty0 215401, 2018{\natexlab{c}}.
\newblock \doi{10.1088/1751-8121/aabcda}.

\bibitem[Bellazzini et~al.(2024)Bellazzini, De~Angelis, and Romano]{Bellazzini:2024dco}
Brando Bellazzini, Stefano De~Angelis, and Marcello Romano.
\newblock {Continuous-Spin Particles, On Shell}.
\newblock arXiv:2406.17017 [hep-th], June 2024.

\bibitem[Metsaev(2025)]{Metsaev:2025qkr}
R.~R. Metsaev.
\newblock {Interacting massive/massless continuous-spin fields and integer-spin fields}.
\newblock arXiv:2406.17017 [hep-th], May 2025.

\bibitem[Brink et~al.(1976)Brink, Deser, Zumino, Vecchia, and Howe]{Brink:1976sz}
L.~Brink, Stanley Deser, B.~Zumino, P.~Di Vecchia, and Paul~S. Howe.
\newblock {Local Supersymmetry for Spinning Particles}.
\newblock \emph{Phys. Lett. B}, 64:\penalty0 435, 1976.
\newblock \doi{10.1016/0370-2693(76)90115-5}.
\newblock [Erratum: Phys.Lett.B 68, 488 (1977)].

\bibitem[Brink et~al.(1977)Brink, Di~Vecchia, and Howe]{Brink:1976uf}
L.~Brink, P.~Di~Vecchia, and Paul~S. Howe.
\newblock {A Lagrangian Formulation of the Classical and Quantum Dynamics of Spinning Particles}.
\newblock \emph{Nucl. Phys. B}, 118:\penalty0 76--94, 1977.
\newblock \doi{10.1016/0550-3213(77)90364-9}.

\bibitem[Berezin and Marinov(1977)]{Berezin:1976eg}
F.~A. Berezin and M.~S. Marinov.
\newblock {Particle Spin Dynamics as the Grassmann Variant of Classical Mechanics}.
\newblock \emph{Annals Phys.}, 104:\penalty0 336, 1977.
\newblock \doi{10.1016/0003-4916(77)90335-9}.

\bibitem[Mondragon et~al.(1994)Mondragon, Nellen, Schmidt, and Schubert]{Mondragon:1994fp}
M.~Mondragon, L.~Nellen, M.~G. Schmidt, and C.~Schubert.
\newblock {Yukawa couplings in the worldline formalism}.
\newblock In \emph{{6th Mexican School of Particles and Fields}}, pages 338--344, 10 1994.

\bibitem[Schuster et~al.(2025)Schuster, Sundaresan, and Toro]{Schuster:2024wjc}
Philip Schuster, Gowri Sundaresan, and Natalia Toro.
\newblock {Thermodynamics of continuous spin photons}.
\newblock \emph{Phys. Rev. D}, 111\penalty0 (5):\penalty0 056019, 2025.
\newblock \doi{10.1103/PhysRevD.111.056019}.

\bibitem[Reilly et~al.(2025)Reilly, Schuster, and Toro]{Reilly:2025lnm}
Aidan Reilly, Philip Schuster, and Natalia Toro.
\newblock {Probing ''Continuous Spin'' QED with Rare Atomic Transitions}.
\newblock arXiv:2505.01500 [hep-ph], May 2025.

\end{thebibliography}








\end{document}